\documentclass[
  ,draft            % you may use draft while you are working on the paper
  ]
  {aipproc}

\layoutstyle{8x11double}

\def\Msun{\ifmmode{\rm M}_\odot\else{M$_\odot$}\fi}
\def\degr{\ifmmode{^\circ}\else$^{\circ}$\fi}

\begin{document}

\title{The Parkes Pulsar Timing Array Project}


\author{R. N. Manchester}{
  address={Australia Telescope National Facility, CSIRO, PO Box 76, Epping NSW 1710, Australia}
}

\begin{abstract}
Detection and study of gravitational waves from astrophysical sources
is a major goal of current astrophysics. Ground-based
laser-interferometer systems such as LIGO and VIRGO are sensitive
to gravitational waves with frequencies of order 100 Hz, whereas
space-based systems such as LISA are sensitive in the millihertz
regime. Precise timing observations of a sample of millisecond pulsars
widely distributed on the sky have the potential to detect
gravitational waves at nanohertz frequencies. Potential sources of
such waves include binary super-massive black holes in the cores of
galaxies, relic radiation from the inflationary era and oscillations
of cosmic strings. The Parkes Pulsar Timing Array (PPTA) is an implementation
of such a system in which 20 millisecond pulsars have been observed
using the Parkes radio telescope at three frequencies at
intervals of two -- three weeks for more than two years. Analysis of these
data has been used to limit the gravitational wave background in our
Galaxy and to constrain some models for its generation. The data
have also been used to investigate fluctuations in the interstellar
and Solar-wind electron density and have the potential to investigate
the stability of terrestrial time standards and the accuracy of
solar-system ephemerides. 
\end{abstract}

%%%%%%%%%%%%%%%%%%%%%%%%%%%%%%%%%%%%%%%%%%%%%%%%%%%%%%%%%%%%%%%%%%%
%%
%% The below \maketitle command inserts the actual front matter data.
%% It has to follow the above declarations.
%%
%%%%%%%%%%%%%%%%%%%%%%%%%%%

\maketitle

\section{Introduction}
The existence of gravitational radiation is a key prediction of
relativistic theories of gravity. These waves propagate at the speed
of light and are generated by the acceleration of massive
bodies. Astrophysical sources of gravitational waves (GW) include
relic radiation from the inflation era \citep{tur97,gri05}, radiation
from reconnection and oscillations of cosmic strings \citep{dv05},
supernovae and formation of compact stars and black holes
\citep{bsr+05}, binary super-massive black holes in the cores of
galaxies \citep{jb03,wl03a,eins04}, coalescence of double-neutron-star
binary systems \citep{kkl+04a} and short-period X-ray binaries
in our Galaxy \citep{nyp04}. Some of these sources may be individually
detectable, others combine to form an essentially isotropic and
stochastic background of GW which permeates all of space. 

Observations of orbital decay of double-neutron-star binary systems
have provided irrefutable evidence that gravitational radiation exists
and that its power is accurately described by Einstein's general
theory of relativity \citep{wt05,ksm+06}. However the signal expected
at the Earth from any realistic source is exceedingly weak, with
typical strain amplitudes of order $10^{-22}$ at frequencies of order
1 Hz. Despite considerable efforts over more than 40 years, up to now
there has been no confirmed direct detection of GW. Initial efforts
used the massive bar detectors pioneered by Joseph Weber
\citep{web69} but more recent detectors with higher
sensitivity are based on laser interferometer systems, for example,
the ground-based systems LIGO \citep{aad+92} and VIRGO \citep{gb02}
and the proposed space interferometer LISA \citep{dan00}.  The
ground-based interferometers are sensitive to GW with frequencies in
the range 10 -- 500 Hz, whereas LISA is sensitive to frequencies in
the range 0.1 -- 100 mHz. Initial LIGO is now operating and has set
limits on various sources \citep[e.g.,][]{aaa+04}; higher sensitivity
will be achieved with Advanced LIGO which is due for completion in
2011. The launch date for LISA is rather uncertain but is unlikely to
be before 2017. 

Pulsars, especially millisecond pulsars (MSPs) are incredibly precise
clocks making possible many interesting applications. Of most interest
to us here is the use of MSPs as GW detectors. GW passing over pulsars
and over the Earth will modulate the received pulsar period; the net
effect is the difference in the modulation at the two ends of the path
\citep{det79}. Pulsar timing experiments measure variations of pulse
phase relative to model predictions. They are therefore most sensitive
to long-period GW with periods comparable to the data span, typically
several years, which corresponds to frequencies in the nanoHertz
regime. Even for these long periods, the expected timing residuals are
very small. Simulations using {\sc tempo2} \citep{hem06,hjl+07} show
that a binary system consisting of two $10^9\;\Msun$ black holes with
a 4-yr orbital period in a galaxy at redshift 0.5 will produce a
timing residual of amplitude just 1.5 ns. Although such a signal would
be very difficult to detect with current technology, expected levels
of the stochastic GW background from binary super-massive black holes
in galaxies are considerably higher with millions of galaxies
throughout the Universe contributing. Predicted levels of the GW
background from other sources such as the inflation era and cosmic
strings, while much more uncertain, are at comparable levels and are
potentially detectable.

Other sources of ``noise'' exist in pulsar timing data and if we wish
to detect GW using pulsar timing we have to be able to separate these
different effects. With timing observations of just one or even a few
pulsars, upper limits may be set but a positive detection is not
possible. However, a large sample of pulsars widely distributed on the
sky --- a pulsar timing array --- can in principle {\em detect} GW with
frequencies in the nanoHertz range.

\section{Pulsar Timing Arrays}
A pulsar timing array consists of a number of pulsars which are widely
distributed on the sky and are observed at (quasi-)regular intervals
over a long time. Typical data spans are many years and typical
observation intervals are a few weeks. To allow correction for
interstellar and Solar-system propagation effects, observations at
several frequencies are required. Such an array has the potential to
make a direct detection of GW with frequencies in the nanoHertz
range. It also can define a ``pulsar timescale'', which may be more
stable over long time intervals than timescales based on atomic
frequency standards, and to detect errors or omissions in the models
of Solar-system dynamics used to define the Solar-system
barycentre. For example, timing arrays have the potential to refine
mass estimates for the outer planets and to detect previously unknown
trans-Neptunian objects. The concept of a pulsar timing array was
first introduced by \citet{hd83} and was further developed (and the
name coined) by \citet{rom89} and \citet{fb90}.

\begin{figure}
\includegraphics[angle=270,width=75mm]{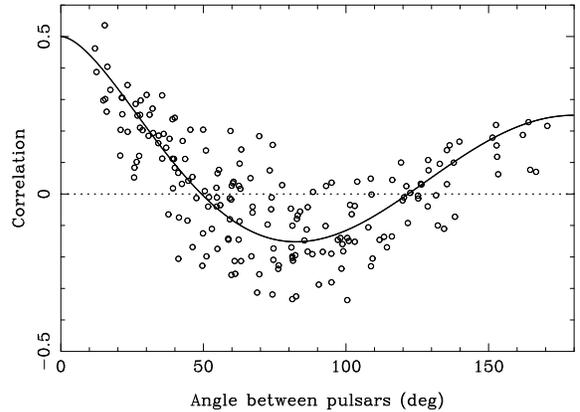}
\caption{Pairwise correlations between simulated timing residuals for
  20 pulsars as a function of angle between the pulsar pairs. The
  timing residuals are dominated by the effects of an isotropic
  stochastic GW background --- there is no intrinsic pulsar period
  noise --- and the scatter of the points around the expected
  correlation results from the stochastic nature of the GW
  signal. \citep{hjl+07}}
\label{fg:gwcorr}
\end{figure}

The key point which enables a pulsar timing array to separate the
effects of a GW from other contributions to observed timing
irregularities is that the signals from the different sources have
different spatial correlation signatures. In other words, the
correlations between timing residuals for pulsars in different
directions on the sky are different for the various noise sources. GW
have quadrupolar symmetry and so pulsars separated on the sky by
$90\degr$ are modulated in opposite senses, whereas those separated by
$180\degr$ have the same sense of modulation. For an isotropic
background, the effect is independent of the orientation of the angle
between the two pulsars. A simulation of the effects of a stochastic
and isotropic GW signal on correlations between timing residuals and
the predicted correlation \citep{hd83} are shown in
Fig.~\ref{fg:gwcorr}.

Although MSP periods are very stable, they are not perfectly
so. Intrinsic timing noise tends to have a red spectrum
\citep{hlk+04}, similar to that of the expected GW
background. However, timing noise in different pulsars is uncorrelated
between the pulsars and so will just add extra noise to the expected
GW signature. Other sources of noise include clock errors and errors
in the planetary ephemeris used to correct observed pulse arrival
times (ToAs) to the Solar-system barycentre. Both of these are
correlated in different pulsars, but they have different spatial
signatures and therefore can be separated. Clock errors will produce
the same residuals for all pulsars and hence will have a spatial
monopole signature. Ephemeris errors are equivalent to an error in the
Earth velocity and hence have a dipole signature on the sky. A
constant error or linear change in clock rate and constant offsets in the
Earth's velocity or acceleration will all be absorbed into the fitted
pulsar periods and period derivatives, but higher order changes in
these quantities are in principle detectable. Likewise, signals from
GW with periods longer than the data span will be absorbed by the
pulsar period fitting. 

\section{The Parkes Pulsar Timing Array}
\begin{figure}
\includegraphics[width=80mm,angle=270]{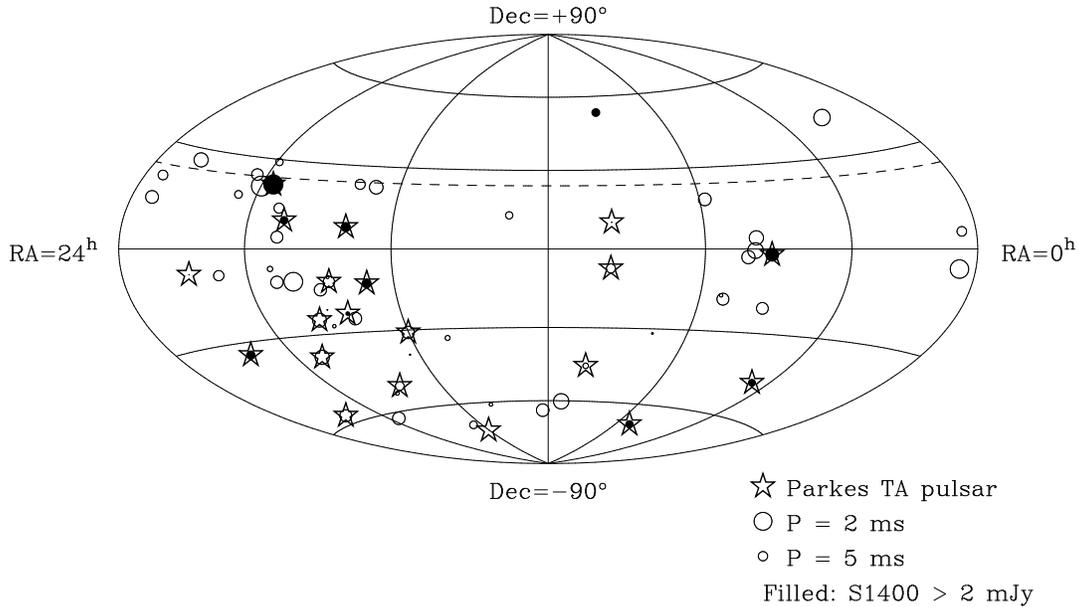}
\caption{Distribution in celestial coordinates of known Galactic disk
  MSPs (with one exception, PSR B1821$-$24 in the globular cluster
  M28) which are suitable for pulsar timing arrays. The size of the
  circle is inversely related to the pulsar period and for stronger
  pulsars the circle is filled. The dashed line is the northern
  declination limit of the Parkes radio telescope. Pulsars being timed
  as part of the PPTA project are marked with stars.}
\label{fg:ppta_psrs}
\end{figure}

The Parkes Pulsar Timing Array (PPTA) project is using the Parkes 64-m
radio telescope to time 20 MSPs at intervals of two to three
weeks. Observations commenced in early 2005 and are made at three
frequencies, 685 MHz (50cm), 1400 MHz (20cm) and 3100 MHz (10cm) with
bandwidths of 64 MHz, 256 MHz and 1024 MHz respectively. The project
is a collaborative effort with principal partners at the Swinburne
University of Technology, the University of Texas at Brownsville and
the ATNF. Fig.~\ref{fg:ppta_psrs} shows the sky distribution of MSPs
which are suitable for pulsar timing array experiments and those
chosen for the PPTA. 

Observations at 1400 MHz normally use the central beam of the Parkes
20cm multibeam receiver \citep{swb+96} which has a system equivalent
flux density of about 30 Jy. A dual-frequency coaxial 10cm/50cm
receiver \citep{gzf+05} allows simultaneous observations at 3100 and
685 MHz; the system equivalent flux densities of these receivers are
approximately 48 Jy and 64 Jy respectively. All receivers receive
orthogonal linear polarisations and have provision for injection of a
linearly polarised calibration signal at $45\degr$ to the two signal
probes. Two main backend systems are used: a digital filterbank (DFB)
and CPSR2, a baseband system allowing coherent dedispersion of two
dual-polarisation 64 MHz bands \citep{hbo06}. The DFB system (PDFB1)
has a maximum bandwidth of 256 MHz and provides on-line correlation
and folding at the topocentric pulsar period giving pulse profiles in
all four Stokes parameters. DFB data files are written using the {\sc
  psrfits} format and all processing uses the {\sc psrchive} data
analysis system \citep{hvm04} and the {\sc tempo2} timing analysis
system \citep{hem06,ehm06,hjl+07}.\footnote{See also
  http://psrchive.sourceforge.net and
  http://www.atnf.csiro.au/research/pulsar/tempo2}

Table~\ref{tb:ppta_psrs} lists the pulsars being observed and gives
the current rms timing residual based on one-hour observations and a
two-year data span. Obvious interference has been excised from the
observations, but they are neither corrected for DM variations nor
accurately calibrated. Only one frequency derivative has been fitted,
apart from PSR J1939+2134 (PSR B1937+21) where three derivatives
are fitted. 

\begin{table}
\caption{PPTA pulsars and their RMS timing residuals}\label{tb:ppta_psrs}
\begin{tabular}{lcccc}
\hline
  \tablehead{1}{c}{b}{PSRJ}
  & \tablehead{1}{c}{b}{Pulse Period\\(ms)} 
  & \tablehead{1}{c}{b}{DM\\(cm$^{-3}$ pc)} 
  & \tablehead{1}{c}{b}{Orbital Period\\(d)} 
  & \tablehead{1}{c}{b}{RMS Residual\\($\mu$s)} \\
\hline
J0437-4715 & 5.757 & 2.65 & 5.74 & 0.12  \\
J0613-0200 & 3.062 & 38.78 & 1.20 & 0.83 \\
J0711-6830 & 5.491 & 18.41 & -- & 1.56 \\
J1022+1001 & 16.453 & 10.25 & 7.81 & 1.11 \\
J1024-0719 & 5.162 & 6.49 & -- & 1.20 \\
J1045-4509 & 7.474 & 58.15 & 4.08 & 1.44 \\
J1600-3053 & 3.598 & 52.19 & 14.34 & 0.35 \\
J1603-7202 & 14.842 & 38.05 & 6.31 & 1.34 \\
J1643-1224 & 4.622 & 62.41 & 147.02 & 2.10 \\
J1713+0747 & 4.570 & 15.99 & 67.83 & 0.19 \\
J1730-2304 & 8.123 & 9.61 & -- & 1.82 \\
J1732-5049 & 5.313 & 56.84 & 5.26 & 2.40 \\
J1744-1134 & 4.075 & 3.14 & -- & 0.65 \\
J1824-2452 & 3.054 & 119.86 & -- & 0.88 \\
J1857+0943 & 5.362 & 13.31 & 12.33 & 2.09 \\
J1909-3744 & 2.947 & 10.39 & 1.53 & 0.22 \\
J1939+2134 & 1.558 & 71.04 & -- & 0.17 \\
J2124-3358 & 4.931 & 4.62 & -- & 2.00 \\
J2129-5721 & 3.726 & 31.85 & 6.63 & 0.91 \\
J2145-0750 & 16.052 & 9.00 & 6.84 & 1.44  \\
\hline
\end{tabular}
\end{table}

The sensitivity of the PPTA to a stochastic background of GW was
investigated by \citet{jhlm05}. They showed that weekly observations
of the 20 MSPs with rms timing residuals of order 100 ns over a
five-year data span was required to detect the predicted GW background
from binary super-massive black holes in galaxies. It is clear that
our current observations do not reach this goal. 

\begin{figure}
\includegraphics[angle=270,width=150mm]{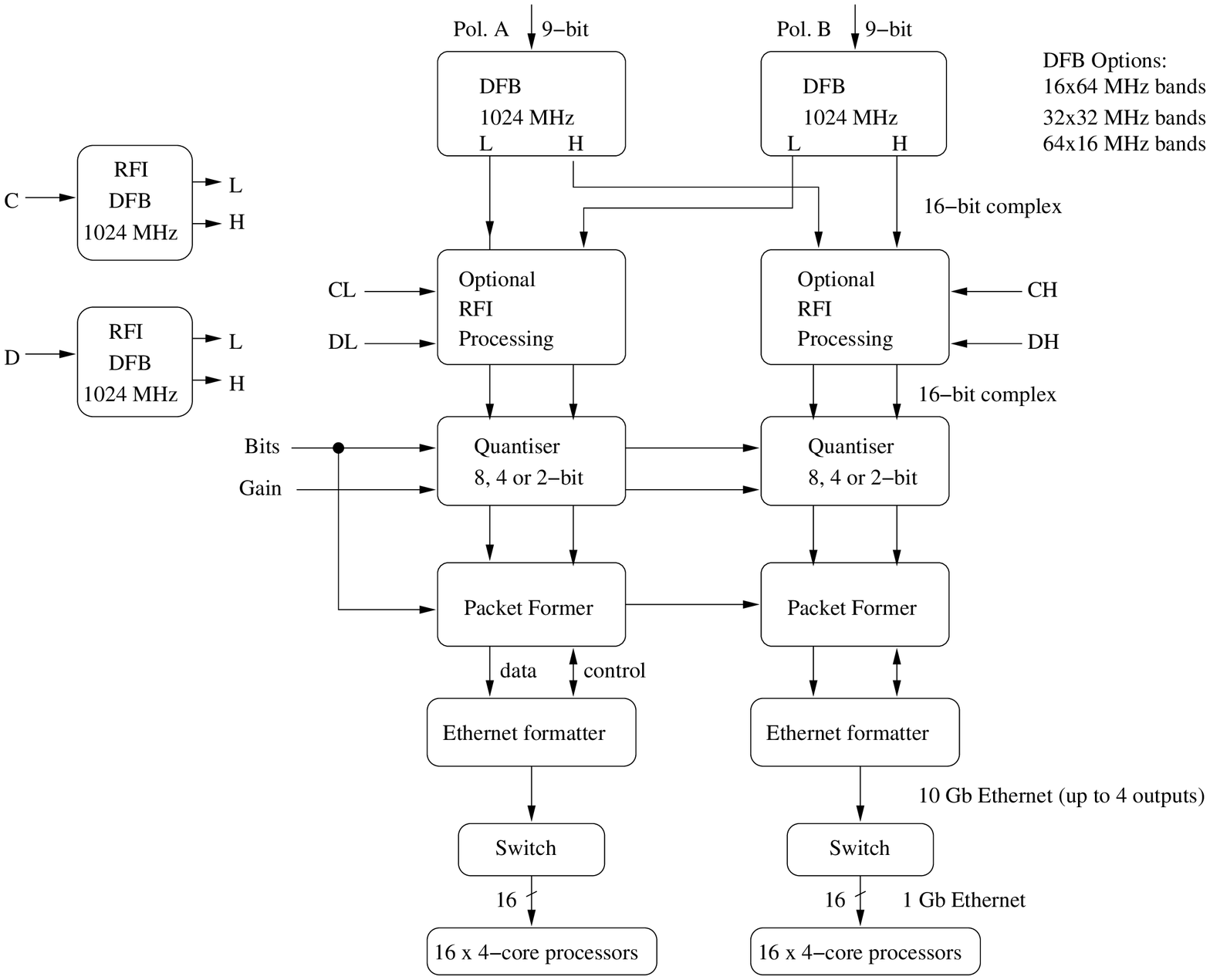}
\caption{Block diagram of the PDFB3/APSR system currently under
  development by the ATNF and Swinburne University. The C and D inputs
may be used for real-time interference mitigation system (as drawn) or
may be used for an independent DFB system. Input signals are Nyquist
sampled with 9-bit precision.}
\label{fg:apsr}
\end{figure}

Several different avenues to improving our data quality are being
explored.  Firstly we are developing new backend systems which will
have better sensitivity and higher time and frequency resolution. A
new DFB system with 1024 MHz bandwidth (PDFB2) was commissioned in
2007 May. First observations with this do have improved performance,
but are still limited for the stronger pulsars by systematic effects
which are not currently fully understood. This system will be further
developed later this year (PDFB3) to double the processing power,
provide real-time mitigation of interference using an adaptive filter
algorithm \citep{khc+05}, write streamed multi-frequency data to disk
for search-mode observations and provide baseband signals over a
maximum bandwidth of 1024 MHz for the next-generation baseband system
(APSR) which is being developed by the Swinburne University group in
conjuction with the ATNF. Fig.~\ref{fg:apsr} shows a block diagram of
the PDFB3/APSR system.

\begin{figure}
\includegraphics[width=130mm]{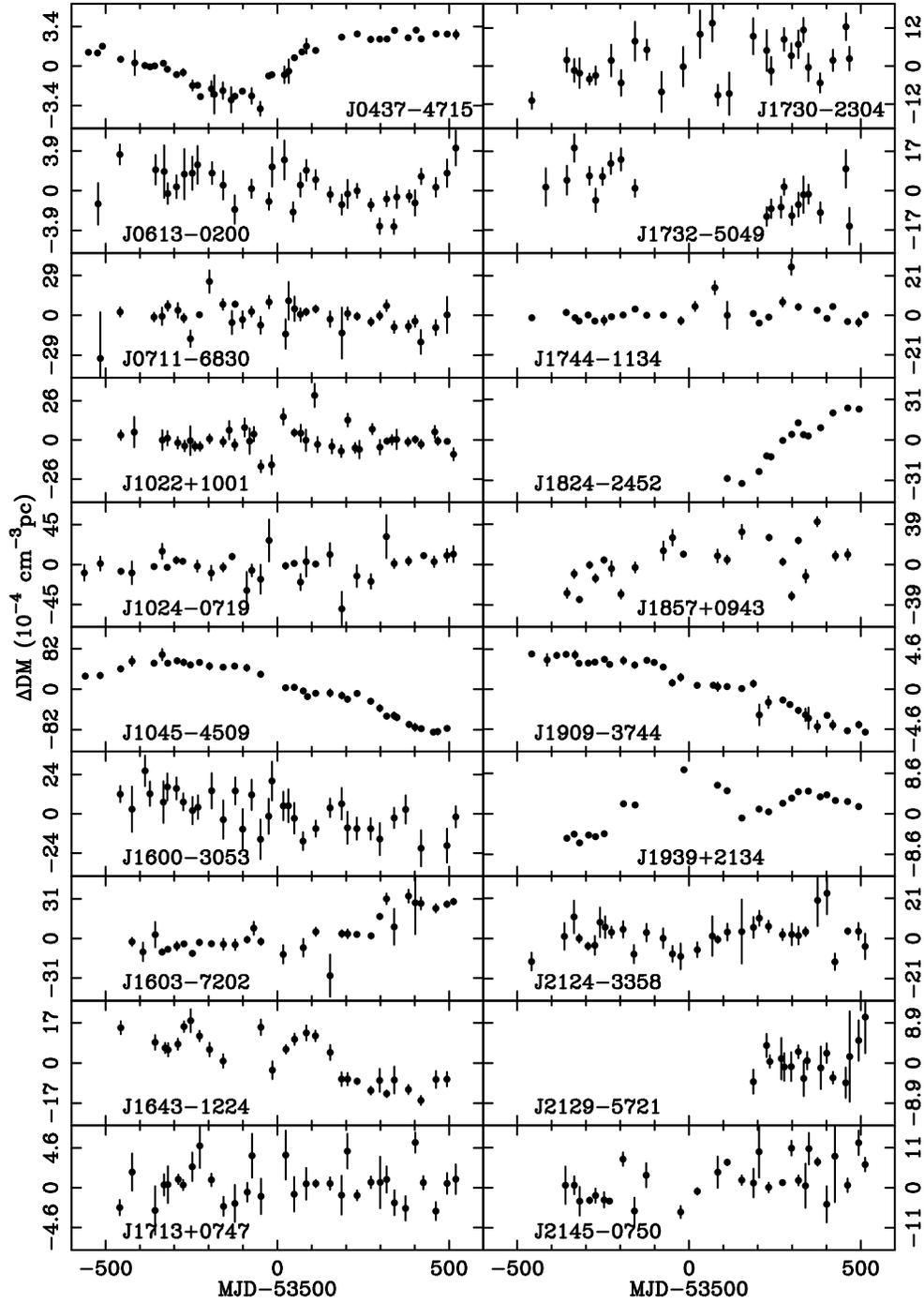}
\caption{Observed variations in dispersion measure for the PPTA pulsars
  obtained by comparing contemporaneous 10cm/50cm or 20cm/50cm ToAs
  \citep{yhc+07}}
\label{fg:dmvar}
\end{figure}

Secondly, we are continually improving our data analysis systems,
{\sc psrchive} and {\sc tempo2}. New methods of displaying, manipulating and
calibrating the data are being developed. As an example, we recently
investigated the effect of DM variations on our data
\citep{yhc+07}. Fig.~\ref{fg:dmvar} shows observed variations in DM
for the 20 pulsars of the PPTA sample. Significant long-term
variations are observed for most of the sample with typical changes of
a few times $10^{-3}$~cm$^{-3}$~pc over the two years. If not taken
into account, these variations will introduce noise into
the measured ToAs making the task of detecting GW, clock errors and
ephemeris errors more difficult. 

\section{Limits on the gravitational-wave background}
Although we do not yet have sufficient sensitivity to detect the
expected stochastic GW background we can put limits on its
amplitude. It is just required that the GW background not contribute a
signal which is detectable in the timing residuals of one or more
pulsars. A 7-yr span of 1400-MHz Arecibo observations of PSR B1855+09
which showed no evidence for low-frequency timing noise was used to
place a limit on $\Omega_{\rm gw}$, the ratio of the energy density of
the GW background in the Galaxy to the closure density of the
Universe, of about $10^{-7}$ \citep{ktr94,mzvl96}. By combining this
data set with PPTA observations of seven pulsars
\citet{jhs+06} were able to reduce this limit
by about an order of magnitude, constraining some models of the relic
GW background and cosmic strings. This result and its implications are
discussed in more detail by Hobbs et al. in these Proceedings. 

\section{A pulsar timescale}
International Atomic Time (TAI) is defined by a weighted average of
many atomic clocks (mostly caesium standards) located at time and
frequency laboratories around the world \citep{pet05}. The most
precise terrestrial timescales available are retroactive revisions to
TAI published by the BIPM \citep{pet03b}, the latest of which is
TT(BIPM06).\footnote{The TT(BIPMxy) timescales may be obtained from
  the Time, Frequency and Gravimetry FTP server at
  http://www.bipm.org.} They differ from TAI by up to several
microseconds and from each other by several tens of nanoseconds,
corresponding to apparent stabilities $\sigma_y \sim 10^{-15}$
\citep{pet05}.

\begin{figure}
\includegraphics[width=80mm,angle=270]{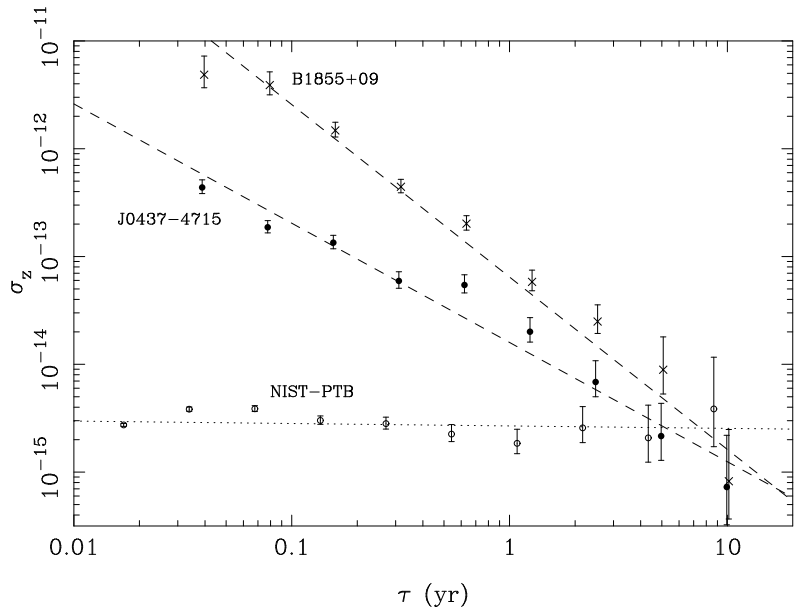}
\caption{Clock stability parameter $\sigma_z$ against averaging time
  for two pulsars, PSR J0437$-$4715 and PSR B1855+09, and for the
  difference between UTC(PTB)
  and UTC(NIST).}
\label{fg:sigmaz}
\end{figure}

A timescale based on pulsars differs fundamentally from these atomic
timescales. First it is based on entirely different physics ---
rotation of massive bodies --- and is largely isolated from
Solar-system and Earth-based effects. Furthermore, pulsars will
continue spinning for billions of years, whereas man-made clocks have
a lifetime measured in years or decades at best. Since current ToA
precisions are at best tens of nanoseconds, stabilities comparable to
those of atomic clocks can only be reached over intervals of several
years. The pulsar timescale is not absolute as it is based on {\it a
  priori} unknown pulsar periods, so the accuracy of the atomic
timescales cannot be checked, only their
stability. Fig.~\ref{fg:sigmaz} shows the stability parameter
$\sigma_z$ \citep{mtem97} for two pulsars with long data spans
together with the same statistic for the difference between two of the
most stable atomic timescales, the German-based UTC(PTB) and the
US-based UTC(NIST). It is clear that, for averaging times of several
years or more, the period stability of MSPs is comparable to and maybe
exceeds that of the best available atomic clocks. Pulsar timing arrays
promise even greater stability as fluctuations in individual pulsars
can be averaged over \citep{rod06}. We can envisage use of a weighted
averaging scheme similar to that applied to the atomic time
standards. Such a pulsar timescale should give better than 50 ns
precision at intervals of weeks and 5 ns precision for fluctuations
with timescales of years, corresponding to a stability of $10^{-16}$
or better.

\section{The future}
The sensitivity of a pulsar timing array to a stochastic GW background
is proportional to the average ToA precision, the square root of the
number of observations (ToAs) and, in the conservative case of no
prewhitening, to the number of pulsars in the array \citep{jhlm05}.
The simulations show that the PPTA alone can just detect the GW
background if it reaches its sensitivity goals. Clearly it is desirable
to increase the number of pulsars observed, the frequency of
observation and the precision of each observation. As discussed above,
we are working on the precision aspect, but it is difficult to
significantly increase the number of pulsars and frequency of
observation for the PPTA. Furthermore, the PPTA is of course limited
to pulsars at declinations south of $+25\degr$, the northern limit of
the Parkes telescope. A wider distribution on the sky would help in
the separation of the different sources of period fluctuation. It
would also be especially valuable for detections of individual sources
of GW, helping to localise them on the sky.

All of these factors point toward the desirability of establishing
international collaborations with other pulsar timing array
projects. We have already established a collaboration with the
European Pulsar Timing Array (EPTA) project \citep{skl+06}, with
agreement on data sharing, coordination of observing schedules and
collaboration on data analysis and interpretation. Future
collaborations with the North American and Chinese pulsar timing
groups, perhaps forming a ``World Pulsar Timing Array'', are under
active discussion.

Looking further to the future, the proposed Square Kilometer Array
(SKA) will have a huge impact on pulsar timing projects. With its
enormous sensitivity and ability to multi-beam, sharing telescope
resources with other projects, sensitive searches and frequent
high-sensitivity observations a large sample of pulsars should be
possible. As an example, let us assume weekly observations of a sample
of 100 MSPs at two or three frequencies with ToA precisions of order
50 ns. Let us further assume that such observations continue for 10
years and take the conservative case of no prewhitening. Provided
intrinsic timing noise in the pulsars does not dominate the effect of
the stochastic GW background, this corresponds to a detection limit at
3 nHz of $\Omega_{\rm gw} \sim 2\times 10^{-13}$. Pre-whitening has the
potential to decrease these limits, but the ultimate level
reached will depend strongly on whether or not MSPs exhibit
significant intrinsic timing noise at these levels of precision over
these long data spans.

\begin{figure}
\includegraphics[width=120mm, angle=270]{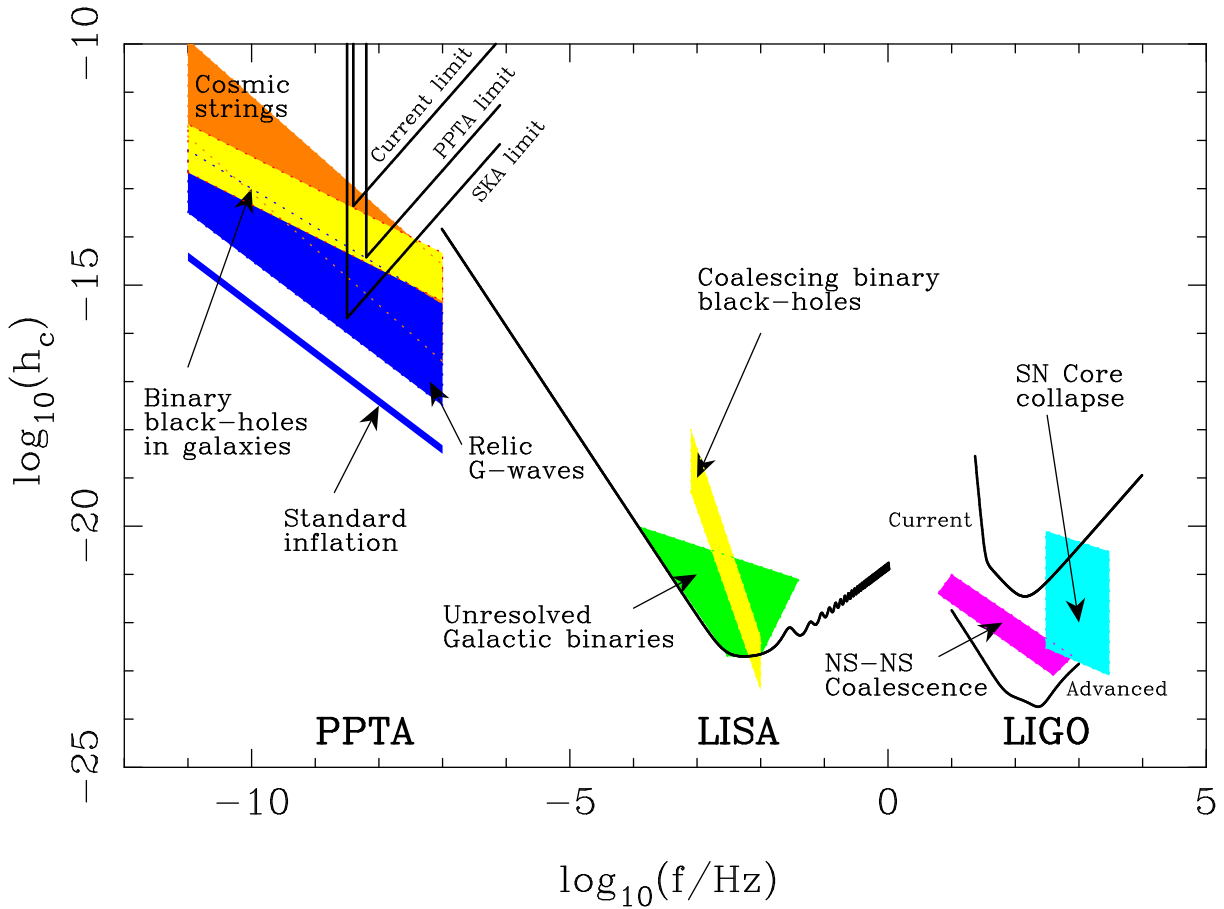}
\caption{Characteristic strain sensitivity for existing and proposed
  GW detectors as a function of GW frequency along with the expected
  levels for signals from some relevant astrophysical sources. }
\label{fg:gwsens}
\end{figure}

Fig.~\ref{fg:gwsens} shows the sensitivity of existing and proposed GW
detectors, illustrating the complementary nature of the different
classes of detector. Expected signal levels from relevant
astrophysical sources in the different frequency bands are also
shown. These cover a range of strain levels. In some cases, for
example, neutron star -- neutron star coalescence, this range reflects
the statistical uncertainty in the occurrence rate of detectable
events, whereas for the sources at nHz frequencies to which the pulsar
timing arrays are sensitive, it more reflects the uncertainty in the
theoretical models on which the predictions are based. For example,
\citet{gri05} predicts a much higher level of relic GW from the
inflationary era than the standard models of inflation
\citep{tur97,bs05}. Sensitivity curves are given for the current
pulsar limits \cite{jhs+06}, the sensitivity goal of the PPTA project
and the 10-year SKA timing array project as described above. With a
15-year data span, the predicted detection limit is approaching the
background level expected from standard inflation. Current
observations are beginning to limit other inflation models and models
for generation of GW by cosmic strings, but do not yet significantly
limit models for formation and evolution of super-massive binary black
holes in the cores of galaxies. The design goal for the PPTA would
either detect this stochastic background or essentially rule out all
current models for its formation. The SKA would give detections with
high significance enabling detailed studies of the source and signal
properties --- an exciting prospect!

\section{Conclusions}
Pulsar timing arrays exploit the remarkable stability of MSP periods
to enable investigation of a range of phenomena. Direct detection of
gravitational waves from astrophysical sources is a major goal of
current astrophysics and pulsar timing arrays have the potential to
achieve this goal. They are sensitive to GW at frequencies of a few
nanoHertz, complementing ground-based and space-based laser
interferometer systems which are sensitive at much higher
frequencies. PTA systems also have the potential to establish a
``pulsar timescale'' which is more stable than the best terrestrial
timescales over intervals of several years or more and to detect
errors or omissions in models of Solar-system dynamics, for example,
the existence of currently unknown trans-Neptunian objects.

The Parkes Pulsar Timing Array (PPTA) project is using the Parkes 64-m
radio telescope to time a sample of 20 millisecond pulsars at three
frequencies every 2 -- 3 weeks. Observations commenced in early 2005,
so we now have over two years of timing data. Sub-microsecond timing
residuals have been achieved on about half the sample but we still
need to improve timing precisions by a factor of a few in order to
have a realistic chance of detecting the stochastic GW background. New
instrumentation and other improvements will help us to achieve that
goal. We are also actively seeking international collaborations with
other timing array projects to increase the sky coverage and the
density of observations. Already, limits on the GW background are
starting to limit some inflation-era and cosmic string models. There
seems little doubt that the proposed Square Kilometer Array radio
telescope will be able to not only detect GW but to also study the
properties of the GW sources in some detail, opening up a new era in
astrophysics. 

%%%%%%%%%%%%%%%%%%%%%%%%%%%%%%%%%%%%%%%%%%%%%%%%
%% BACKMATTER
%%%%%%%%%%%%%%%%%%%%%%%%%%%%%%%%%%%%%%%%%%%%%%%%

\begin{theacknowledgments}
The Parkes Pulsar Timing Array project has a great team of people based
at the collaborating institutions. They contribute in different ways to
helping us achieve our goals and I thank them all for their efforts. The
Parkes radio telescope is part of the Australia Telescope which is
funded by the Commonwealth of Australia for operation as a National
Facility managed by CSIRO.
\end{theacknowledgments}

%\bibliographystyle{aipproc}   % if natbib is available
%\bibliography{journals,modrefs,psrrefs,crossrefs}

\end{document}